\documentclass[a4paper,prl,twocolumn,showpacs]{revtex4}

\usepackage{graphicx}
\usepackage{subfigure}
\usepackage{dcolumn}
\usepackage{bm}

\def\bra#1{\mathinner{\langle{#1}|}}
\def\ket#1{\mathinner{|{#1}\rangle}}

\begin{document}

\title{Graphene as a reversible spin manipulator of molecular magnets}
\author{Sumanta Bhandary$^{\dagger}$, Saurabh Ghosh$^{\ddagger}$, Heike Herper$^{\star}$, Heiko Wende$^{\star}$, Olle Eriksson$^{\dagger}$ and Biplab Sanyal$^{\dagger}$ \cite{bs}}
\affiliation{$^{\dagger}$ Department of Physics and Astronomy, Uppsala University, Box 516,
 751\,20 Uppsala, Sweden}
\affiliation {$^{\ddagger}$ School of Applied and Engineering Physics, Cornell University, Ithaca New York 14853 USA}
\affiliation{$^{\star}$ Faculty of Physics and Center for Nanointegration Duisburg-Essen (CeNIDE),\\ University of Duisburg-Essen, Lotharstr. 1, 47048 Duisburg, Germany}
 
\date{\today}

\begin{abstract}
One of the primary objectives in molecular nano-spintronics is to manipulate the spin states of organic molecules with a d-electron center, by suitable external means. 
In this letter, we demonstrate by first principles density functional calculations, as well as second order perturbation thoery, that a strain induced change of the spin state, from S=1 $\to$ S=2, takes place for an iron porphyrin (FeP) molecule deposited at a divacancy site in a graphene lattice. The process is reversible in a sense that the application of tensile or compressive strains in the graphene lattice can stabilize FeP in different spin states, each with a unique saturation moment and easy axis orientation. The effect is brought about by a change in Fe-N bond length in FeP, which influences the molecular level diagram as well as the interaction between the C atoms of the graphene layer and the molecular orbitals of FeP.

\end{abstract}


\maketitle





Fe-porphyrin (FeP) is a potential candidate among metalloporphyrin molecules for biomolecular devices to be used for spin dependent electronic transport using the characteristics of spin switching. A significant effort has been made to switch the properties of organic molecules by the underlying magnetic substrates. By a combined study of element-specific magnetic measurements and first principles electronic structure calculations, it was shown that it is possible to tune the magnetic coupling between Fe at the center of FeP with a ferromagnetic substrate like Co or Ni \cite{heikonatmat, heikoprl}. Also, switching properties of molecules can be driven by some external means like temperature, pressure, light and external ligands. It should be noted that the change in the spin state of the FeP molecule depends on the ligated or non-ligated situation \cite{fepligand}. A ligated FeP molecule consisting of 5 Fe-3d electrons, can have only odd moments (S= 1/2, 3/2, 5/2) whereas a non-ligated one, with 6 Fe-3d electrons, has even spin states (S= 0, 1, 2). Recently, it was shown that the spin of metalloporphyrins exchange coupled with ferromagnetic films, can be controlled by nitric oxide molecule acting as an axial ligand \cite{organo}. In a theoretical work \cite{cho}, an organometallic framework based on Cr-porphyrin arrays was proposed as a spin filter in spintronic devices.

Graphene has made enormous sensation in the scientific community due to its exotic multidisciplinary features \cite{graphene}, which have the potential to be exploited in future electronics, gas sensing, batteries, understanding of high energy experiments etc. During the synthesis of graphene or at a post-synthesis step involving functionalization with e.g. an acid treatment \cite{coleman}, a number of defects may arise. In sp$^2$ bonded carbon, the divacancy defect can easily be introduced in a graphene lattice \cite{defects} and these are the defects we propose to utilize here. It has been shown that the electronic and transport properties of graphene are affected significantly due to the presence of divacancy defects \cite{jafri,sanyalprb}. The chemisorption energies at these defect sites are quite large and hence, divacancy defects behave as excellent trapping centers for adatoms and molecules. \cite{sanyalprb, arkadyprl}.

Utilizing the novel properties of FeP and graphene, in this letter, we propose an alternative method to manipulate the spin states of a FeP molecule placed at a divacancy center in a graphene lattice. Our aim of the current study is to investigate the chemical and magnetic interactions of FeP with a defected graphene lattice and to tune the magnetic properties of the molecule by modifying its electronic and magnetic states by strain engineering of graphene.  We have also studied the modification of electronic states in graphene due to the presence of FeP. To this end, we have performed state-of-the-art first principles calculations that show that the strain engineering of graphene lattice modifies the spin state of FeP in a reversible way.

We have performed first principles density functional calculation using the VASP code \cite{vasp}. Plane wave projector augmented wave basis was used in the local spin density approximation (LSDA) for the exchange correlation potential. The plane wave cut off energy used was 400 eV. A 3x3x1 Monkhorst Pack k-points set was used for the integration in Brillouin zone. Atoms were relaxed until the Hellman Feynman forces were minimized up to 0.01 eV/\AA. For the FeP-graphene composite system, we have used a 8x8 supercell of graphene with 21 \AA~of vacuum length perpendicular to the graphene plane in the unit cell. To account for the narrow d-states of Fe in the FeP molecule, we have used a LSDA+U approach where the strong Coulomb interaction is added according to the mean field Hubbard U formalism. The values of Coulomb parameter U and exchange parameter J used are 4 eV and 1 eV respectively as these values reproduce correctly the electronic structure and magnetism of FeP in the gas phase \cite{chemphys}.

\begin{figure}
\begin{center}
\includegraphics[scale=0.35]{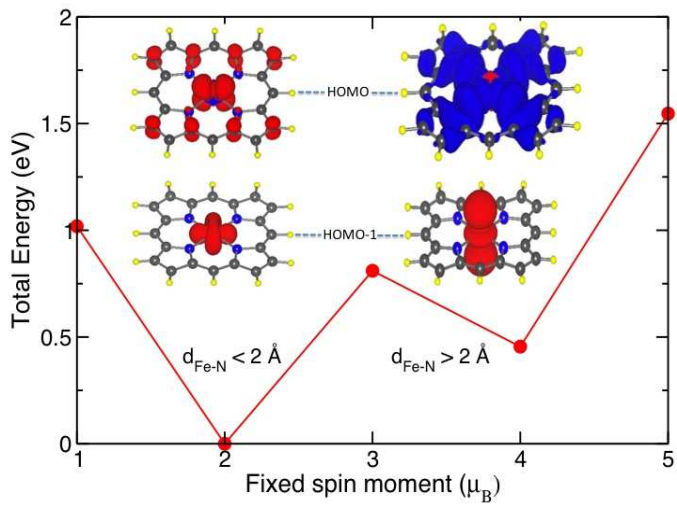}
\caption{(Color online) Total energies (shown with respect to the ground state) of FeP in the gas phase, plotted as a function of fixed spin moments. The charge densities for HOMO and HOMO-1 energy levels are shown for 2 $\mu_{B}$ and 4 $\mu_{B}$ cases. Red and blue colors indicate spin-up and spin-down electron density, respectively. In the FeP molecule, Fe, N, C and H atoms are drawn as red, blue, gray and yellow balls, respectively.} 
\label{fig1}
\end{center}
\end{figure}

Let us first discuss the spin states of FeP in the gas phase. We have 
performed fixed spin moment calculations for the FeP molecule and the geometry of the molecule is allowed to  relax to the ground state. Our calculations reveal that apart from the ground state minimum of the total energy with a magnetic moment of 2 $\mu_B$, another minimum at 0.55 eV higher energy than the ground state, exists with a moment of 4 $\mu_B$, as seen in Fig.~\ref{fig1}. The energy barrier between the 2 $\mu_B$ and 3 $\mu_B$ magnetic states is found to be 0.81 eV, indicating that one has to overcome at least this amount of energy to switch the spin state from 2 $\mu_B$ to 4 $\mu_B$. We will below refer to these states as the intermediate spin state (S=1) and high spin state (S=2), respectively. The ground state minimum occurs for a Fe-N bond distance of 1.96 \AA~ whereas the higher energy minimum has a Fe-N bond at 2.03 \AA~. Hence, the spin-state is intimately coupled to the Fe-N bond distance. We also show the magnetization densities corresponding to Highest Occupied Molecular Orbital (HOMO) and HOMO-1 energy levels for the 2 $\mu_B$ and 4 $\mu_B$ cases. For the S=1 spin state, the dominant contributions to HOMO and HOMO-1 levels arise from d$_{\pi}$ and d$_{xy}$ orbitals of Fe, respectively, whereas  d$_{x^2-y^2}$ and d$_{z^2}$ orbitals of Fe are relevant for the S=2 spin state. So, a clear change of orbital symmetry accompanying the change of the spin state is observed.
  
In our calculations for the FeP-graphene composite system, FeP is initially placed in such a way that the center Fe atom of the FeP molecule is on the top of the divacancy center  and the geometry as well as the spin state are optimized to achieve the ground state. The calculated binding energy (B.E) is 0.28 eV, which demonstrates that the system is chemically stable. In the optimized geometry, the vertical separation between FeP and graphene becomes 3.2 \AA.  The local geometry of the molecule remains
more or less the same  as in the gas phase, i.e the Fe atom almost remains in the molecular plane with a vertical deviation of 0.05 \AA~ from the molecular plane. The Fe-N bond 
distance is found to be 1.96 \AA~ (same as in the gas phase) with a magnetic moment of 2.18 $\mu_B$. The small difference in magnetic moment of the free molecule with an integer moment of 2 $\mu_B$ is due to the interaction of the molecule with the divacancy site, which makes the edge C atoms to be spin polarized. 

\begin{figure}
\begin{center}
\includegraphics[scale=0.37]{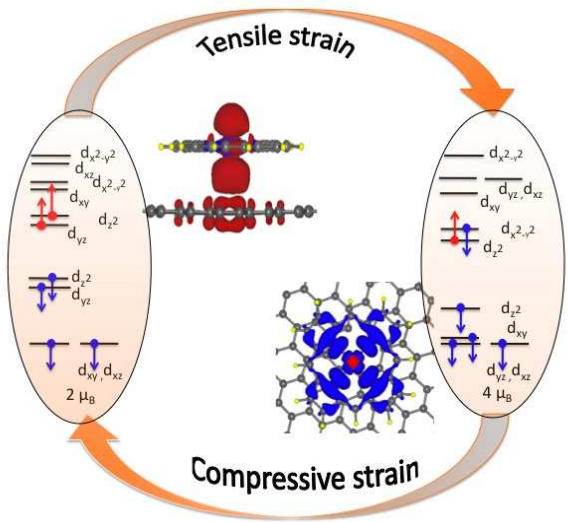}
\caption{(Color online) Magnetization density isosurfaces for FeP on (left) 0\% and (right) 1\% strained graphene. The isosurfaces have been plotted for an energy window of 0.4 eV below the Fermi levels in both cases. The spin-up and spin-down densities are drawn in red and blue, respectively. The energy levels with the d-orbital character of FeP are shown in the extreme left and right for 0\% and 1\% strained graphene, respectively. } 
\label{fig2}
\end{center}
\end{figure}

In Fig.~\ref{fig2},  we show the orbital occupancies of Fe-3d states in FeP when trapped at the divacancy defect site, in the 0\% and 1 \% strain situations. 
For free FeP with a total magnetic moment 2 $\mu_B$, the d$_{xy}$ orbital is completely filled, the d$_{z^2}$ is half filled with one spin-down electron and the d$_{yz}$ and d$_{xz}$ orbitals are degenerate in energy having two spin-down and one spin-up electron. In case of FeP trapped at a divacancy defect site in the 0\% strain situation, the d$_{xy}$ orbital becomes half filled with one spin-down electron, the degeneracy between d$_{yz}$  and d$_{xz}$ is lifted, d$_{yz}$ becomes completely filled whereas d$_{xz}$ becomes partially filled with one spin-down electron. The most interesting change is found in the d$_{z^2}$ orbital. The unoccupied part of the spin-up d$_{z^2}$ orbital of the Fe atom of free FeP, becomes partially occupied when adsorbed. This is clearly observed in the magnetization density isosurface plot in Fig.~2. The change in the orbital arrangement is due to the interaction between FeP and the divacancy defect of graphene leading to a non integer magnetic moment of the combined system. 

We now address the possibility to rearrange the orbital occupation of the Fe-d levels and hence to modify the spin state of the system by strain engineering of graphene with divacancy defects. We have exerted tensile strains by increasing the unit cell vectors of the unstrained graphene lattice, considering the cases of 1\% and 2\% strained graphene with FeP placed at the top of the defect site and have performed geometry optimizations. In case of 1\% strain, the B.E of the FeP molecule is found to be 0.45 eV. Unlike the case of FeP on unstrained graphene, FeP on strained graphene goes through a structural deformation due to an enhanced interaction with the divacancy center. In the unstrained situation, the  divacancy defect reconstructs to a 5-8-5  structure and a clear $\pi$-band formation is observed (data not shown). But once the size of the vacancy site is increased due to the tensile strain, the $p_{z}$  orbitals of the edge C atoms do not take part in the formation of a $\pi$-band. Instead they form more localized levels, with unsaturated bonds. The structural rearrangement increases the Fe-N bond length from 1.96~\AA~to 2.03~\AA~and this leads to a change in spin state from S=1 to S=2 of FeP. The energy barrier of 0.81 eV between these two spin states, observed in the gas phase, is overcome due to the interaction between FeP and the strained graphene lattice. It is important to note that the spin crossover behavior of molecules on surfaces is hardly realized in practice due to the strong interaction between the two systems. \cite{gutlich}. On the contrary, in this work, we demonstrate that it is the surface-molecule interaction that helps in the spin transition.

It is interesting to investigate hysteresis effects and whether a reversible process is possible via a compressive strain which may cause a transition from the S=2 state to the S=1 state. If possible, this could lead to an ultrathin spintronic device having logic states assigned as '0' and '1', depending on the spin state (S=1 or S=2), which may be influenced by strain effects. To test this hypothesis, we released the 1\% strain and exposed the FeP-graphene entity to a 2\% compressive strain. For this situation, the geometry optimized structure has the Fe-N bond length equal to 1.96 \AA~ with an accompanying transition from the high spin (S=2) to the intermediate spin (S=1) state. Hence, a situation of reversible spin switching can be achieved by the application of a relatively small tensile or compressive strain, as illustrated also in Fig.~\ref{fig2}.

To analyze the situation further we note that in the 1\% tensile strained condition, the d$_{z^2}$ orbital becomes completely filled (which is significantly different from the unstrained case), the d$_{xy}$ orbital is partially filled (one spin-down electron), the d$_{yz}$ and d$_{xz}$ orbitals become degenerate in energy and are partially filled (with two spin-down electrons). More importantly, the d$_{x^2-y^2}$ orbital, which is completely empty for the ground state of FeP in the gas phase as well as on the 0\% strained graphene, becomes partially occupied, with this orbital being close to the Fermi level, as clearly seen in the magnetization density isosurface plot in Fig.~2. The orbital occupations are similar to those of gas phase FeP in the high spin state. This indicates that the interaction between FeP and  the divacancy in a strained graphene lattice mainly overcomes the barrier between S=1 and S=2 spin states in realizing a high spin ground state of FeP. The application of a compressive strain (2\%) brings back FeP to the S=1 spin state with the orbital arrangement of FeP similar to a 0\% strained case. It should also be mentioned that the role of a divacancy in spin switching is crucial. We have found from our calculations that without this defect, the change in spin state of FeP is not possible when a strain is applied to graphene.

Experimentally, the straining of the graphene lattice can be realized by straining the substrate on which graphene is deposited and the predicted reversible change in the spin state may be achieved, which can be monitored by the x-ray magnetic circular dichroism (XMCD) measurements of electronic structure of Fe 3d states \cite{stohr}. The discussion of the spin moments of FeP measured by XMCD requires special attention as the contribution of the spin-dipole term $\langle T_{z} \rangle $ is known to be quite large \cite{gambardella} for this class of organic molecules deposited on substrates. Therefore, we have calculated the $7 \langle T_{z} \rangle$  values following the method prescribed by Wu and Freeman.\cite{tz} The values are -1.89 $\mu_{B}$ and 1.32 $\mu_{B}$ for 0$\%$ strained (S=1) and 1$\%$ strained (S=2) graphene lattice respectively. The absolute values are quite large and are similar in magnitude to what have been observed by Stepanow {\it et al.} for CuPc. \cite{gambardella} One interesting point to note is the change in sign of $7 \langle T_{z} \rangle$ due to the change in the spin state and hence, change in the occupation of d-orbitals of different symmetries. This dramatic change will strongly affect the effective spin moment (2$S_{z}+ 7 \langle T_{z} \rangle$) measured by XMCD. 

\begin{figure}
\begin{center}
\includegraphics[scale=0.4]{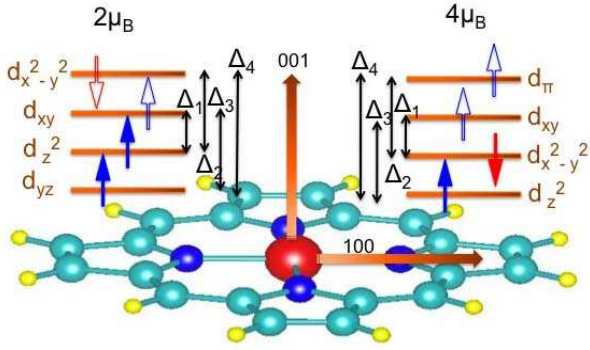}
\caption{(Color online) Schematic diagram showing the energy levels and their orbital characters for (left) 0$\%$ and (right) 1$\%$ strained cases, considered for the calculation of MAE (see text). The filled and empty arrows indicate occupied (HOMO and HOMO-1) and unoccupied (LUMO and LUMO+1) d-states of Fe, respectively. The big arrows indicate the directions of magnetization. $\Delta$s are defined in the text.} 
\label{fig3}
\end{center}
\end{figure}
The consequences due to the change in the spin state of FeP on graphene under strain are expected to be manifested in the uniaxial magnetic anisotropy too. To study this, we have considered the effect of spin-orbit (SO) interaction through the hamiltonian $H_{SO}=\zeta {\bf L.S}$, where $\zeta$ is the SO coupling constant with ${\bf L}$ and ${\bf S}$ being the orbital and spin angular momenta respectively. According to second order perturbation theory, the SO contribution to the energy is : $E_{SO}=-\zeta^2\sum_{u,o}\frac{| \bra{u}{\bf L.S}\ket{o}|^{2}}{E_{u}-E_{o}}$. Here $\ket{o}$ and $\ket{u}$ correspond to occupied and unoccupied states, respectively, with $\ket{u/ o}$=$\ket{lm,\sigma}$, where $l$, $m$ and $\sigma$ are azimuthal, magnetic and  spin quantum numbers, respectively. Such an approach was e.g. used in Ref. \onlinecite{zeta}.
We have considered four relevant states in this calculation, HOMO-1, HOMO, LUMO, LUMO+1, to estimate the SO contribution. All other states would cause a too large denominator in the expression of the MAE, and hence were neglected. We start by defining the following matrix elements: $\Delta_1 = \bra{d_{xy},\alpha}{\bf L.S}\ket{d_{z^2},\alpha}; \Delta_2 = \bra{d_{xy},\alpha}{\bf L.S}\ket{d_{yz},\alpha};  \Delta_3 = \bra{d_{x^2-y^2},\beta}{\bf L.S}\ket{d_{z^2},\alpha}; \Delta_4 = \bra{d_{x^2-y^2},\beta}{\bf L.S}\ket{d_{yz},\alpha}$ as depicted in Fig.~\ref{fig3} with $\alpha$ and $\beta$ representing spin-up and spin-down states. For the magnetization along the z-axis, we get  $E^{001}_{SO}=-\sum_{i}\frac{|\Delta_i|^2}{\delta_iE}$, where $\Delta_1=0, \Delta_2=0, \Delta_3=0, \Delta_4=\frac{i}{2}$ and $\delta_iE=E_{u}-E_{o}$. Hence, $E^{001}_{SO}=-\frac{\zeta^2}{4(E_{d_{x^2-y^2}}-E_{d_{yz}})}$.

When the magnetization is along the x-axis, the spin functions of the local coordinate system can be expressed in terms of global spinor functions as: $ \chi^{\uparrow}=\frac{1}{\sqrt{2}}(\alpha+\beta)$ and  $\chi^{\downarrow}=\frac{1}{\sqrt{2}}(\alpha-\beta)$. With a similar approach as above, we obtain $\Delta_1 = \Delta_2 = \Delta_3 = \Delta_4=0$ and hence, $E^{100}_{SO}=0$. Therefore $MAE=E^{001}_{SO}- E^{100}_{SO}=-\frac{\zeta^2}{4(E_{d_{x^2-y^2}}-E_{d_{yz}})}$, with the 001 axis as the easy axis. In a similar fashion, for 1$\%$ strained graphene with S=2, we  obtain~$MAE=[\frac{\zeta^2}{(E_{d_{xy}}-E_{d_{x^2-y^2}})}+\frac{\zeta^2}{4(E_{d_{xz}}-E_{d_{x^2-y^2}})}+\frac{3\zeta^2}{4(E_{d_{yz}}-E_{d_{z^2}})}]-\frac{\zeta^2}{2(E_{d_{xz}}-E_{d_{x^2-y^2}})}$, where we identify HOMO-1 as $\ket{d_{z^2},\alpha}$, HOMO as $\ket{d_{x^2-y^2},\beta}$, LUMO as $\ket{d_{xy},\alpha}$ and LUMO+1 as $\ket{d_{\pi},\alpha}$. One should note that LUMO+1 is now a doubly degenerate $d_{\pi}$ state, so we need to consider five states to calculate the $MAE$.

For  $MAE$ calculations, $\zeta$ for Fe in a 2+ state, has been taken as 0.05 eV. \cite{zeta} The final results are the following: {\it for 0$\%$ strain (S=1 of FeP), MAE = -0.4 meV and the 001 axis is the easy axis, whereas for 1$\%$ strain, S=2 and the MAE is 2 meV, with the easy axis being in the plane of the molecule}. This is a quite interesting result as due to the change in the spin state, the easy axis switches from out-of-plane to in-plane direction with a 5-fold increase in the absolute value of MAE.

In conclusion, we propose here a new direction of combining  molecular electronics and graphene science to achieve novel switching properties. By first principles calculations, we have shown that the S=1 magnetic state of FeP trapped at a divacancy site of a graphene lattice can be switched to a S=2 spin state by applying a tensile strain in the defected graphene lattice. One may think of other alternatives for the substrate, e.g., polymers, which can be stretched too. However, the effect discussed here relies on a strong binding between the FeP and the substrate through the defect site, so that any substrate strain translates to the FeP molecule. For graphene with divacancies this is indeed the case, as shown here, and it remains to be seen if it also holds for other substrates.
The doubling of the saturation moment represents a big magnetoelastic effect, which is found to be reversible since a compressive strain can bring back the S=1 spin state from a S=2 spin state. The microscopic mechanism of the spin switching is the change in Fe-N bond length of FeP and the concomittant change in the electronic structure and the level arrangement of molecular orbitals that hybridize with defect induced states in graphene. We also show that the modification of the spin state is associated with a modification of the out-of-plane easy axis to an in-plane one, which is detectable, e.g., in XMCD measurements. 

We gratefully acknowledge financial support from STINT {\it Institutional Grant for Young Researchers}, Swedish Research Council (VR) and a KOF grant from Uppsala University. O.E. is in addition grateful to the KAW foundation and the ERC (project 247062 - ASD) for support. We also acknowledge SNIC-UPPMAX, SNIC-HPC2N and SNIC-NSC centers for the allocation of time in high performance supercomputers.

\end{document}